\documentclass[final]{article}

\usepackage[nonatbib]{neurips_2023}

\usepackage[utf8]{inputenc} 
\usepackage[T1]{fontenc}    
\usepackage{hyperref}       
\usepackage{url}            
\usepackage{booktabs}       
\usepackage{amsfonts}       
\usepackage{nicefrac}       
\usepackage{microtype}      
\usepackage{xcolor}         
\usepackage{graphicx}
\usepackage{enumitem}

\title{JungleGPT: Designing and Optimizing Compound AI Systems for E-Commerce}

\author{%
Sherry Ruan \\
  Stanford University \\
  \texttt{sruan@cs.stanford.edu} \\
  \And
  Tian Zhao
    \\
  Stanford University\\
  \texttt{tianzhao@cs.stanford.edu} \\
}

\begin{document}

\maketitle

\begin{abstract}
  LLMs have significantly advanced the e-commerce industry by powering applications such as personalized recommendations and customer service. However, most current efforts focus solely on monolithic LLMs and fall short in addressing the complexity and scale of real-world e-commerce scenarios. In this work, we present JungleGPT, the first compound AI system tailored for real-world e-commerce applications. We outline the system's design and the techniques used to optimize its performance for practical use cases, which have proven to reduce inference costs to less than 1\% of what they would be with a powerful, monolithic LLM. 
\end{abstract}

\section{Introduction}

Global e-commerce sales are projected to reach \$6.3 trillion in 2024 \cite{emarketer}, accounting for 6\% of the global gross domestic product (GDP) \cite{visualcapitalist, imf}, underscoring the critical role of e-commerce in the world economy.
AI has been instrumental in accelerating the e-commerce industry by enhancing the online shopping experience through personalized recommendations and facilitating online sales with AI-powered tools such as creative content generation and automated customer service.

With recent advancements in large language models (LLMs), significant progress has been made in e-commerce, including e-commerce-specific translations \cite{General2Specialized}, product review analysis \cite{ROUMELIOTIS2024100056}, and automatic product descriptions \cite{zhou2023leveraging}. 
Furthermore, researchers have employed various LLM techniques to enhance the performance of LLMs on specific e-commerce tasks, including crafting instruction datasets tailored for e-commerce \cite{peng2024ecellm}, instruction tuning \cite{Li_Ma_Wang_Huang_Jiang_Zheng_Xie_Huang_Jiang_2024}, and multi-aspect instruction following \cite{shi2023llamae}. These methods have proven to surpass the capabilities of general LLMs within the e-commerce domain. However, existing work primarily focuses on improving the performance of monolithic LLMs on e-commerce datasets. Real-world e-commerce user patterns differ significantly from traditional enterprise AI in three key areas:
\begin{itemize}[leftmargin=*]

\item{\textbf{Global Footprint}: Both e-commerce buyers and sellers are located globally, which requires \textit{efficient caching systems} to ensure seamless access to data.}

\item \textbf{Read-Heavy Operations}: E-commerce user patterns generally involve more read operations than write operations. Therefore, it is essential to bring snapshots of data closer to users, such as \textit{within browser sessions or on edge networks}.

\item \textbf{Long-Tail Users}: Online sellers are predominantly small to medium-sized businesses (SMBs) (e.g., about 98\% of Amazon sellers are SMBs \cite{gitnux}). The ratio of their data size to their paying power is typically high, so LLM solutions need to be \textit{extremely cost-efficient}.
\end{itemize}

\section{JungleGPT: An E-Commerce AI Compound System}

To address the limitations of monolithic LLMs described above, we designed and built \textit{JungleGPT}, the first compound AI system \cite{compound-ai-blog} tailored for e-commerce to meet the unique needs of this industry.

\paragraph{Design} 

\begin{figure}
    \centering
    \includegraphics[width=\textwidth]{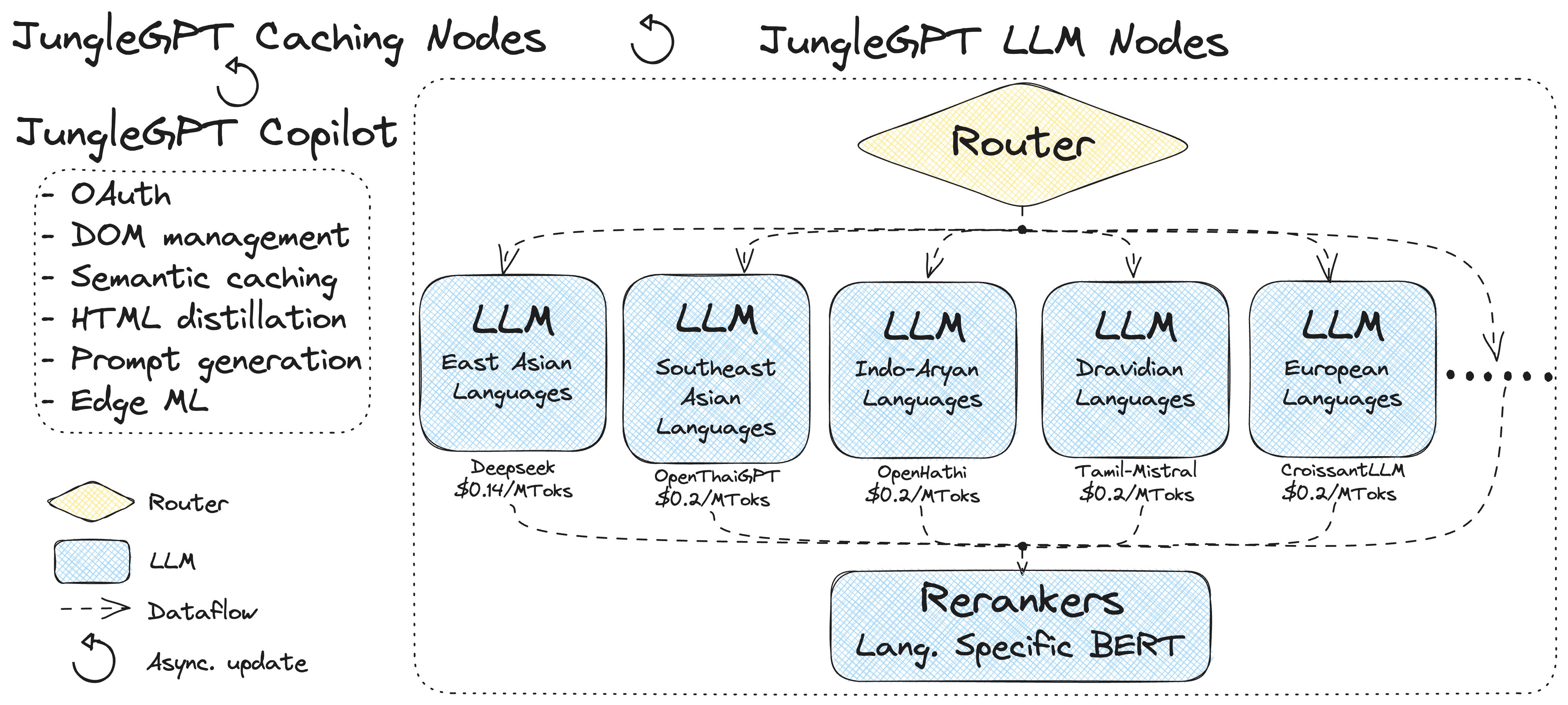}
    \caption{JungleGPT Compound AI System Design}
    \label{fig:junglegpt-system}
\end{figure}
The JungleGPT system encompasses three key components: \textit{JungleGPT Copilot}, \textit{JungleGPT Caching Nodes}, and \textit{JungleGPT LLM Nodes}. The components are connected with asynchronous updates to ensure prompt user interactions.
\begin{itemize}[leftmargin=*]
    \item \textbf{JungleGPT Copilot}: JungleGPT Copilot sits on the critical path of user interaction loops. Conceptually, this is the ``L1 cache'' of the whole JungleGPT system - data storage and compute capacity are low and designed to support frequent read operations. The copilot runs fast, lightweight machine learning models using modern web technology, e.g., webGPU \cite{kenwright2022introduction} for analyzing user queries and returning responses with low latency. Furthermore, JungleGPT Copilot utilizes powerful parsers in modern web frameworks to clean up context for LLMs and prepare prompts. To ensure security, we design semantic caches specifically for sensitive user data, e.g., personal identification information. We implement these caches using browser-local session memories and do not share them with backend JungleGPT LLM nodes.
    
    \item \textbf{JungleGPT Caching Nodes}: E-commerce users' geographical locations are extremely global. To enable low-latency, distributed data access for users, we utilize modern caching systems deployed on edge networks - each user can access a snapshot of their data on edge nodes that are geographically closer to them, and these caches are updated lazily by the backend JungleGPT LLM nodes. In our evaluation, we typically find users able to access their snapshots within hundreds of milliseconds. 
    
    \item \textbf{JungleGPT LLM Nodes}: JungleGPT LLM nodes are detached from the main user interaction loop and periodically update JungleGPT Caching Nodes. The asynchronous design ensures that LLM inference latency does not sit on the critical path in user interaction.
\end{itemize}

\paragraph{Cost Optimization}
    Cost efficiency is a key consideration for e-commerce applications. Generally, e-commerce SaaS products have monthly fees capped at several hundred dollars. Assuming a computing budget of \$100/month, a user can only generate 2.2M tokens with GPT4.5 \cite{OpenAI_Pricing}, which significantly falls short of compute needs. From our user interviews, we typically see users trying to analyze texts that are 1 to 2 orders of magnitude greater than 2.2M tokens. Furthermore, e-commerce users are often English-as-a-Second-Language (ESL) speakers and require LLMs to provide language-specific analysis. To meet these needs, we utilize an ensemble of small, cost-efficient LLMs \cite{CroissantLLM, OpenHathi, OpenThaiGPT, bi2024deepseek, Tamil-Mistral-7B-Instruct-v0.1} fine-tuned on common languages of our user base. To further improve analysis quality, we create lightweight rerankers fine-tuned for each language group. Combining small LLMs and rerankers fine-tuned for non-English use cases reduces our inference cost to less than $1\%$ of using a powerful and monolithic LLM endpoint.

\paragraph{Conclusion}
We identify three quadrants: global footprint, read-heavy operations, and long-tail users where current monolithic LLMs for e-commerce fall short and a compound AI system can bring significant value to end users. JungleGPT, a compound AI system for e-commerce, is designed to address system requirements in these three quadrants. We hope this work inspires more research and industry efforts towards building effective compound AI to advance the e-commerce industry.

{
\bibliographystyle{plain}
\bibliography{sample}


\end{document}